\documentclass[preprint]{revtex4-2}  
\usepackage{graphicx}
\usepackage[utf8]{inputenc}
\usepackage{lineno}
\usepackage{amsmath}      

\begin{document}          

\title{Prediction of the treatment effect of FLASH radiotherapy with Circular Electron-Positron Collider (CEPC) synchrotron radiation}

\author{Junyu Zhang}
\address{The Institute for Advanced Studies, Wuhan University, Wuhan, China}
\author{Xiangyu Wu}
\address{The Institute for Advanced Studies, Wuhan University, Wuhan, China}
\author{Pengyuan Qi}
\email{qpyuan@hust.edu.cn}
\address{Cancer Center, Union Hospital, Tongji Medical College, Huazhong University of Science and Technology, Wuhan, China}
\author{Jike Wang}
\email{Jike.Wang@whu.edu.cn}
\address{The Institute for Advanced Studies, Wuhan University, Wuhan, China}

\date{\today}
	
\begin{abstract}
The Circular Electron-Positron Collider (CEPC) can also work as a powerful and excellent synchrotron light source, which can generate high-quality synchrotron radiation. This synchrotron radiation has potential advantages in the medical field, with a broad spectrum, with energies ranging from visible light to x-rays used in conventional radiotherapy, up to several MeV. FLASH radiotherapy is one of the most advanced radiotherapy modalities. It is a radiotherapy method that uses ultra-high dose rate irradiation to achieve the treatment dose in an instant; the ultra-high dose rate used is generally greater than 40 Gy/s, and this type of radiotherapy can protect normal tissues well. In this paper, the treatment effect of CEPC synchrotron radiation for FLASH radiotherapy was evaluated by simulation. First, Geant4 simulation was used to build a synchrotron radiation radiotherapy beamline station, and then the dose rate that CEPC can produce was calculated. Then, a physicochemical model of radiotherapy response kinetics was established, and a large number of radiotherapy experimental data were comprehensively used to fit and determine the functional relationship between the treatment effect, dose rate and dose. Finally, the macroscopic treatment effect of FLASH radiotherapy was predicted using CEPC synchrotron radiation light through the dose rate and the above-mentioned functional relationship. The results show that CEPC synchrotron radiation beam is one of the best beams for FLASH radiotherapy.
\end{abstract}
\maketitle
	
\section{Introduction}
	
The Circular Electron-Positron Collider (CEPC)\cite{cepc2018cepc} is a large-scale international scientific project initiated and hosted by China. It is an electron positron collider with a circumference of 100 kilometers, including linear accelerators, energy intensifiers, colliders and other important components. CEPC can also work as a powerful and excellent synchrotron light source. It has electrons with energy up to 120 GeV which is much higher than that of any other synchrotron light source and can produce better quality synchrotron radiation when electrons move around the storage ring. The light source has a wide spectrum, from visible light to x-rays (several hundred keV) used in conventional treatments, reaching up to several MeV.
	
At present, FLASH radiotherapy is one of the hot research topics in the medical field. However, the research of photon FLASH radiotherapy lags behind that of electron FLASH radiotherapy. This is because compared with the more easily obtained electron linear accelerator, the generation of X-rays in photon linear accelerator is limited by electron heat deposition. The number of synchrotron radiation facilities available is relatively small.\cite{montay2022flash} Therefore, photon FLASH radiotherapy equipment is also one of the research directions. Photon linear accelerators suitable for FLASH radiotherapy have been successfully designed,\cite{liu2023development} and synchrotron radiation which can provide higher dose rates is also worthy of attention. The excellent properties and spectrum of CEPC are of great help to the frontier research in the medical field, so it makes sense to study the feasibility of using CEPC synchrotron radiation for advanced radiation therapy FLASH radiotherapy.
	
FLASH radiotherapy is a radiotherapy method that uses ultra-high dose rate irradiation to instantly reach the treatment dose.\cite{favaudon2014ultrahigh}\cite{maxim2019flash} The ultra-high dose rate used is generally greater than 40 Gy/s. FLASH radiotherapy can protect normal tissue by a mechanism known as the FLASH effect. When FLASH effect occurs, normal tissue shows reduced toxicity while tumor control is unchanged, so while killing tumor cells, normal tissue can be effectively protected to avoid adverse reactions. Therefore, FLASH radiotherapy is a better treatment method than conventional radiotherapy, which has become a research hotspot in the field of radiotherapy.
	
Some studies have shown that FLASH radiotherapy requires the delivery of pulsed rays in an extremely short time, which is a necessary condition. FLASH radiotherapy has great advantages over traditional radiotherapy. One drawback of traditional radiotherapy is that the beam can cause great damage to the normal tissue around the tumor, which can affect human health. FLASH radiotherapy does not reduce the killing ability of tumor tissue while protecting normal tissue. In addition, the movement of tumor cells due to breathing during radiotherapy can affect the accuracy of beam irradiation, reduce the effectiveness of treatment and cause greater damage to the surrounding normal tissue. FLASH radiotherapy, due to the extremely short time, can greatly improve the treatment effect and avoid the motion error in the treatment process.\cite{wang2021flash} The advantages of FLASH radiotherapy have been found in studies of multiple tissues and organs, such as the lungs, brain, skin, intestines, and blood.\cite{borghini2024flash} It has even shown good treatment effects on human skin.\cite{bourhis2019treatment}
	
The mechanism of FLASH radiotherapy is one of the hot topics in current research, and many explanations have been put forward.\cite{gao2022potential} First, in terms of oxygen depletion, the ultra-fast delivery of single-dose FLASH radiotherapy depletes oxygen in normal tissues, thereby increasing their resistance to radiation.\cite{montay2019long}\cite{pawelke2021electron}\cite{pratx2019ultra} It seems that oxygen depletion at least partially explains how FLASH radiotherapy protects normal tissue from damage. For tumors, local oxygen depletion induced by FLASH radiotherapy has little impact, whose radiation response is decided by the total dose delivered. Second, a large number of reactive oxygen species can be generated in a very short time at the ultra-high dose rate, and the dense reactive oxygen species undergo self-recombination and transform into substances harmless to normal cells, resulting in FLASH effect.\cite{labarbe2020physicochemical}\cite{hu2023radical}\cite{wardman2020radiotherapy} Third, for immune response, FLASH radiotherapy will reduce the number and time of irradiation of immune cells and reduce the damage to the immune system.\cite{jin2020ultra}\cite{venkatesulu2019ultra}
	
Although there are many explanations, the mechanism of FLASH radiotherapy remains unclear. Therefore, it is difficult to directly combine all mechanisms to model and evaluate the therapeutic effect of FLASH radiotherapy on CEPC synchrotron radiation. Rudi Labarbe et al. solved the 9 differential rate equations resulting from the radiolytic and enzymatic reactions network using the published values of these reactions rate constants in a cellular environment and proposed a physicochemical model of reaction kinetics to explain normal tissue sparing by FLASH radiotherapy.\cite{labarbe2020physicochemical} This model built the relationship between dose rate, dose and the normal tissue complications probability (NTCP), which could be used to estimate roughly the treatment effect of medical beams. Therefore, we built a synchrotron radiation beamline using Geant4 simulation software and the treatment effect of FLASH radiotherapy with CEPC synchrotron radiation was predicted with the model above.
	
\section{Methods}
	
\subsection{Calculation of CEPC bending magnet source photon number}

The energy of electrons in the CEPC storage ring is 120 GeV. Synchrotron radiation is produced when electrons pass through bending magnet. The total energy emitted by a single electron per turn is
\begin{equation}
U=88.46 \times \frac{E^{4}}{\rho}
\end{equation}
where the unit of radiated energy $U$ is keV, the unit of electron energy $E$ is GeV, and the unit of curvature radius $\rho$ is m.

If the current of the circulating particles is $I$, the total power of synchrotron radiation is 
\begin{equation}
P=88.46 \times \frac{E^{4} \times I}{\rho}
\end{equation}

where the unit of power $P$ is W and the unit of current $I$ is mA.

So the radiated power per unit length is 
\begin{equation}
P_{l}=88.46 \times \frac{E^{4} \times I}{\rho} \div 2\pi\rho = 14.08 \times \frac{E^{4} \times I}{\rho^{2}}
\end{equation}

The critical energy of synchrotron radiation is used to characterize the ``hardness" of radiation. The emitted radiation energy on both sides of the critical energy is equal. It is defined by the following expression
\begin{equation}
E_{c}=2.218 \times \frac{E^{3}}{\rho}
\end{equation}

The photon spectrum produced by a single electron is
\begin{equation}
\frac{d^{2}N}{d \varepsilon dt}=(\frac{2\alpha}{\sqrt{3}h})(\frac{1}{\gamma^{2}})\int_{\gamma}^{\infty}K_{\frac{5}{3}}(\eta)d\eta
\end{equation}

where $\alpha$ is the fine structure constant, $h$ is Planck's constant, $\gamma$ is the ratio of total energy $E$ to $m_{0}c^{2}$ and $K$ is the modified Bessel function order $ \frac{5}{3} $.

Finally, the total number of photons per second per unit length is calculated to be $2.363\times 10^{16}$ photons/m/s.

For CEPC, the bending angle of bending magnet is 2.844 mm$\cdot$rad and the radius of curvature is 10 700 m, so the number of photons emitted by bending magnet is $7.19\times 10^{17}$ photons/s.
Therefore, we can estimate the dose rate of CEPC medical beamline by using Monte Carlo simulation to get the average dose produced by a single photon.

\subsection{Simulation method}
This article uses two simulation tools: SHADOW and Geant4.10.2. SHADOW software is used to generate photons emitted by the bending magnet with the Monte Carlo method.\cite{lai1986shadow}\cite{sanchez2011shadow3} The CEPC and bending magnet parameters including the electron beam spot size, horizontal and vertical divergence and energy are shown in Table 1\cite{cepc2018cepc}, which were input for simulation to obtain the phase space state of photons, including position, direction, energy and polarization.

\begin{table}
	\caption{CEPC and bending magnet parameters}
	\begin{tabular}{lcr}    
		Parameter & Value & Unit       \\
		\hline
		energy & 120 & GeV \\
		$\sigma_{x}$ & $2.09 \times 10^{-2}$ & mm \\
		$\sigma_{y}$ & $6.8 \times 10^{-5}$ & mm \\
		$\varepsilon_{x}$ & 1.21 & nm$\cdot$rad \\
		$\varepsilon_{y}$ & $3.1 \times 10^{-3}$ & nm$\cdot$rad \\
		radius of curvature & 10700 & m \\
		\hline
	\end{tabular}
\end{table}

The process of photon transmission from bending magnet to the patient's position was simulated by Monte Carlo method using Geant4.\cite{agostinelli2003geant4} The layout of the beamline (see Fig. 1) is simplified without complex elements and divided into two cases, one without low-energy filters and the other with filters. In the figure, SAD refers to the source-to-aim distance. The setting of the distance of each component from the source point is inspired by the layout of the ESRF ID17 beamline. The low-energy filters are composed of carbon, aluminum and copper foils with the thickness of 1.42 mm, 1.52 mm and 1.04 mm, respectively, which also refer to the ones in ESRF ID17 beamline. The patient is replaced by the water phantom which is used to measure the deposited energy of the beam. And a 1 mm × 1 mm slit is added 1.7 m in front of the water phantom, which limits the size of the field. The tungsten is selected as the material of the slit due to its high density, which can effectively block out-of-field photons. In addition, the slit thickness is set at 50 cm in our simulations. While this thickness is unrealistic for practical applications, it was intentionally chosen to be large enough to block all photons, thereby ensuring it is sufficient for achieving the desired outcomes in our model.

\begin{figure}
	\includegraphics[width=1\textwidth]{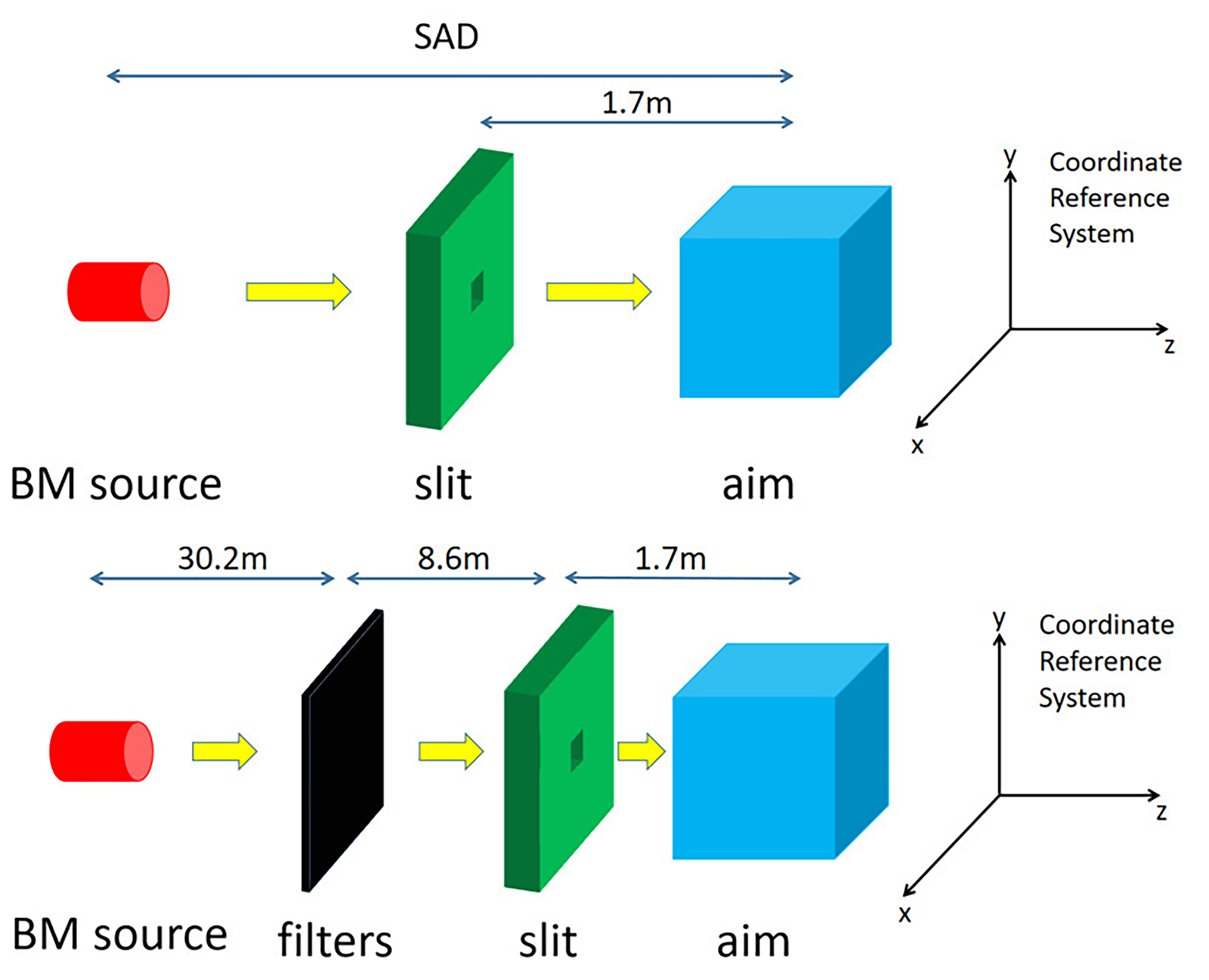}
	\caption{The simplified layout of CEPC beamline elements from the bending magnet to the aim position for simulation.}
\end{figure}

With the help of two simulation tools, we can simply evaluate the quality of the CEPC synchrotron radiation beam and calculate the dose rate of CEPC under the simple beamline.

\subsection{Prediction model of FLASH radiotherapy}
The occurrence of radiotherapy involves many physical and chemical reactions, such as the radiolysis of water, reactions with oxygen and radical reactions with bio-molecules.\cite{spitz2019integrated} Rudi Labarbe et al. solved the 9 differential rate equations resulting from the radiolytic and enzymatic reactions network using the published values of these reactions rate constants in a cellular environment and proposed a physicochemical model of reaction kinetics to explain normal tissue sparing by FLASH radiotherapy.\cite{labarbe2020physicochemical} The 9 differential rate equations involve nine substances including $e_{aq}^{-}$, $O_{2}$, $H_{2}O_{2}$, $OH\cdot$, $H\cdot$, $H_{2}$, $O_{2}\cdot ^{-}$, $R\cdot$, and $ROO\cdot$, which are described in Table 2. 

\begin{table}
	\caption{The description of nine substances.}
	\begin{tabular}{ll}    
		\hline
		$e_{aq}^{-}$ & Electrons in aqueous solution, generated from water radiolysis. \\
		$O_{2}$ & Molecular oxygen, naturally present in tissues. \\
		$H_{2}O_{2}$ & Hydrogen peroxide, produced further from radiolysis products interacting with oxygen. \\
		$OH\cdot$ & Hydroxyl radicals, a product of water radiolysis. \\
		$H\cdot$ & Hydrogen atoms, a product of water radiolysis. \\
		$H_{2}$ & Molecular hydrogen, formed from water radiolysis. \\
		$O_{2}\cdot ^{-}$ & Superoxide anion, formed from the reaction of radiolysis products with oxygen. \\
		$R\cdot$ & Free radicals derived from the radiolytic decomposition of carbon-based biomolecules (RH). \\
		$ROO\cdot$ & Peroxyl radicals, harmful to lipids and DNA, formed from the combination of $R\cdot$ with oxygen. \\
		\hline
	\end{tabular}
\end{table}     

Peroxyl radical $ROO\cdot$ is considered to be the main cause of harmful effects on lipids and DNA, and the production of $ROO\cdot$ can be considered to be related to the degree of cell damage. Therefore, for semi-quantitative predictions, it is speculated that the cell biological response and the NTCP are a sigmoid function. Based on the existing experimental data, the parameters in the function are fitted. 

This model establishes the relationship between dose rate, dose and treatment effect well. So we will use this model to roughly predict the treatment effect of FLASH radiotherapy under CEPC synchrotron radiation.

\section{Results and discussion}

\subsection{Source characteristic}
The photons emitted by the bending magnet was generated by SHADOW software.
The parameters of CEPC and bending magnet were input for simulation to obtain the phase space state of photons, including energy, position, direction, and so on. The photon distribution is shown in Fig. 2, in which Figure 2a is the cross-sectional spatial distribution of the beam; Figures 2b and 2c are the relationship between position and direction of photons. These figures show that the distribution of photons produced by SHADOW is axially symmetrical.

\begin{figure}
	\includegraphics[width=1\textwidth]{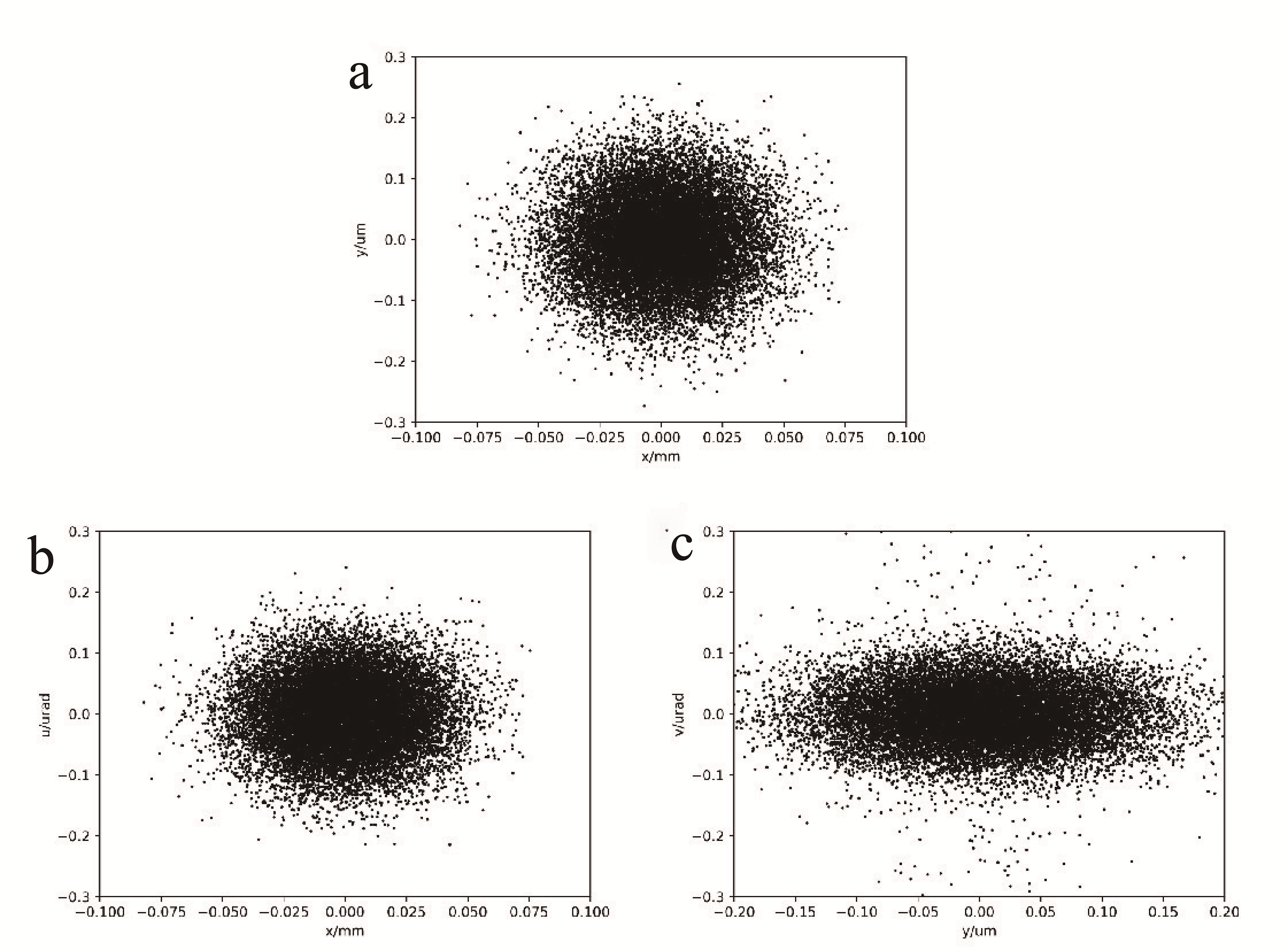}
	\caption{Beam cross section at the the bending magnet exit (a). Phase-space diagrams in the horizontal (b) and vertical (c) coordinates at the end of the bending magnet. Each point represents a photon.}
\end{figure}

The simulation results of photon number distribution along the X and Y axis are shown in Fig. 3, where the maximum number of photons is normalized to 1. The position where the photon information is collected is 35 cm in front of the water phantom. The number distribution of photons is flat, and there is no large fluctuation. Without filters, the average energy of the beam is 134 keV. What's more, the average energy of the beam with filters increases to 307 keV, which is better than 100 keV of ESRF ID17 medical beamline. Energy affects the depth of treatment; the higher the energy, the deeper the photon enters into the human tissue.

\begin{figure}
	\includegraphics[width=1\textwidth]{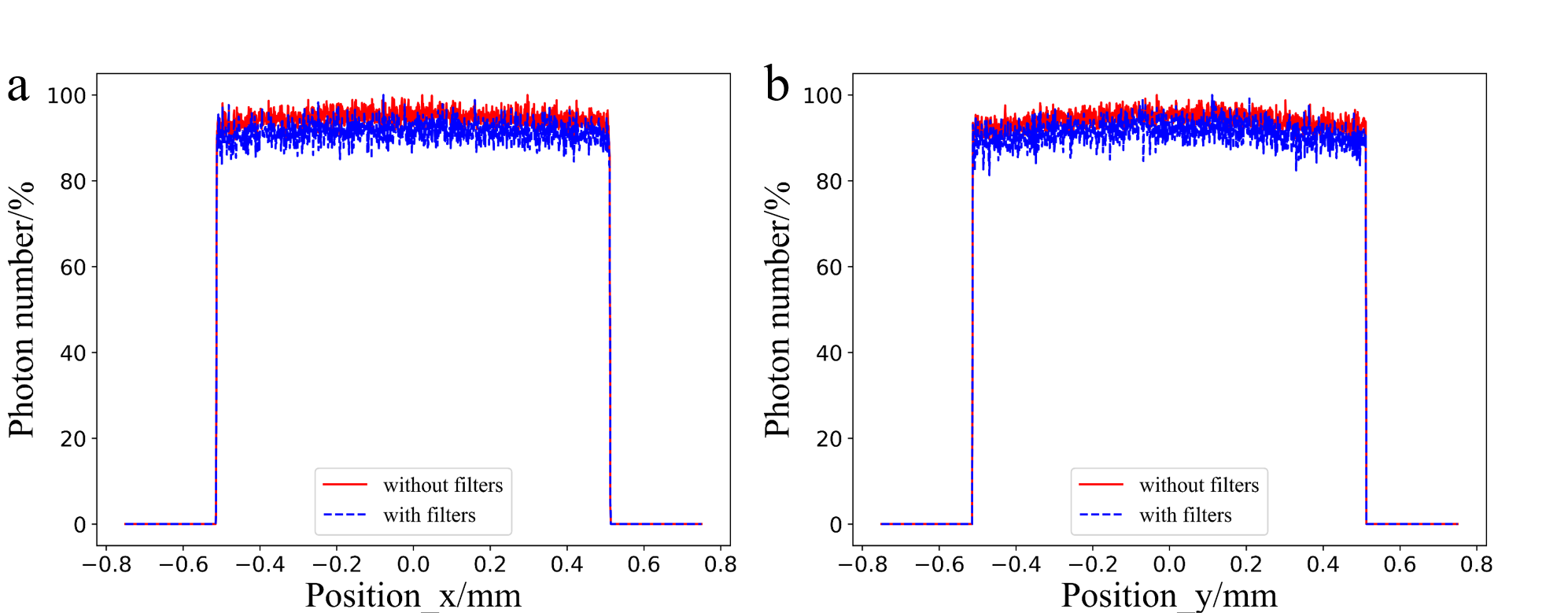}
	\caption{The number distribution of photons which is 35 cm in front of the water phantom in which the maximum number of photons is normalized to 1.}
\end{figure}

\subsection{Dose calculation}
In radiotherapy, percentage depth dose (PDD) is commonly used to characterize the treatment depth. PDD relates the absorbed dose deposited by a radiation beam into a medium as it varies with depth along the axis of the beam. The dose values are divided by the maximum dose, referred to as $d_{max}$, yielding a plot in terms of percentage of $d_{max}$.
It can be seen from the Fig. 4a that the treatment depth of the beam with filters is much greater than that of the beam without filters. The PDD curves for both a 9MeV electron beam and a 9MeV photon beam(general LINAC; TrueBeam model of VARIAN) using Geant4, without filters, were simulated and compared to the CEPC beam. The result demonstrates that the CEPC beam (average energy is 307 keV with filters) can offer a slightly better penetration depth than the 9 MeV electron LINAC. By optimizing the filter structure, the average energy of the beam can be increased and its penetration depth will be improved. While the penetration depth of a 9 MeV photon LINAC is superior to that of the CEPC beam, it is important to note that the dose rate from a general LINAC is quite low. For example, the TrueBeam medical LINAC, made by VARIAN, can deliver a dose rate of about 10 Gy/min. In contrast, the CEPC synchrotron beam can deliver a very high dose rate, which is a significant advantage for certain therapeutic applications.

We selected 85\% of $d_{max}$ as the cutoff point, above which the effective treatment range was defined. It can be seen that the treatment depth of the beam without filters is only several millimetres, while the treatment depth of the beam with filters can reach around 2 cm, which is enough for treating superficial tumors. While the CEPC beam is currently not suitable for treating deep-seated tumors, its effective treatment depth of 2 cm is superior to other synchrotron radiation sources. In addition, further optimizations in filter structure and increasing the average energy of the beam could potentially enhance the effective treatment depth. These kinds of optimizations would be the next focus of our work.

Dose profile is used to characterize the horizontal dose distribution, where the maximum dose without filters is normalized to 1. As can be seen from the Fig. 4b, the dose distribution is symmetric. Within the slit range, the doses at each position which are between 85\% and 100\% of the maximum dose are considered to be effective. This uniform distribution of the dose across the entire irradiation field means that the beam is suitable for radiotherapy.

\begin{figure}
	\includegraphics[width=1\textwidth]{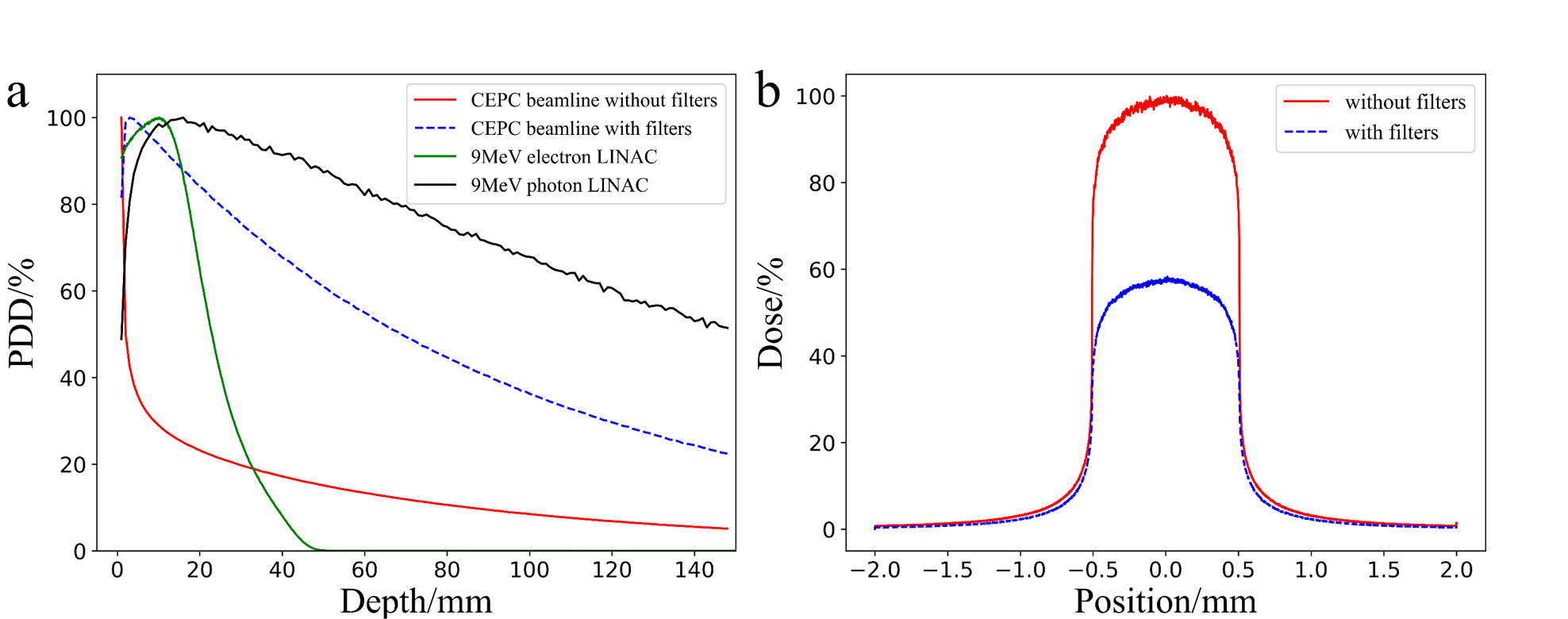}
	\caption{Dose distribution. (a) Percentage depth dose with different conditions in which the maximum dose is normalized to 1. (b) Dose profile in the horizontal direction in which the maximum dose without filters is normalized to 1.}
\end{figure}

Polarization leads to an anisotropic lack of photons compared to non-polarized isotropic scattering. This can lead to differences in the dose calculation at the slit edge with and without considering polarization. In clinical treatment, too many differences will affect the treatment plan and effect. The effect of polarization on dose calculation is also the case for the CEPC synchrotron beam. Fig. 5 shows the differences at the edge of the slit between dose calculations regarding and ignoring photon polarization without filters. The width of the slit is 1 mm, so the -0.5 mm and 0.5 mm positions in the horizontal direction are the focus of attention, and the differences are small, less than $2\%$. In addition, the speed of the simulation calculation can be improved by not considering the polarization. These advantages are due to the high quality of the CEPC beam.

\begin{figure}
	\includegraphics[width=1\textwidth]{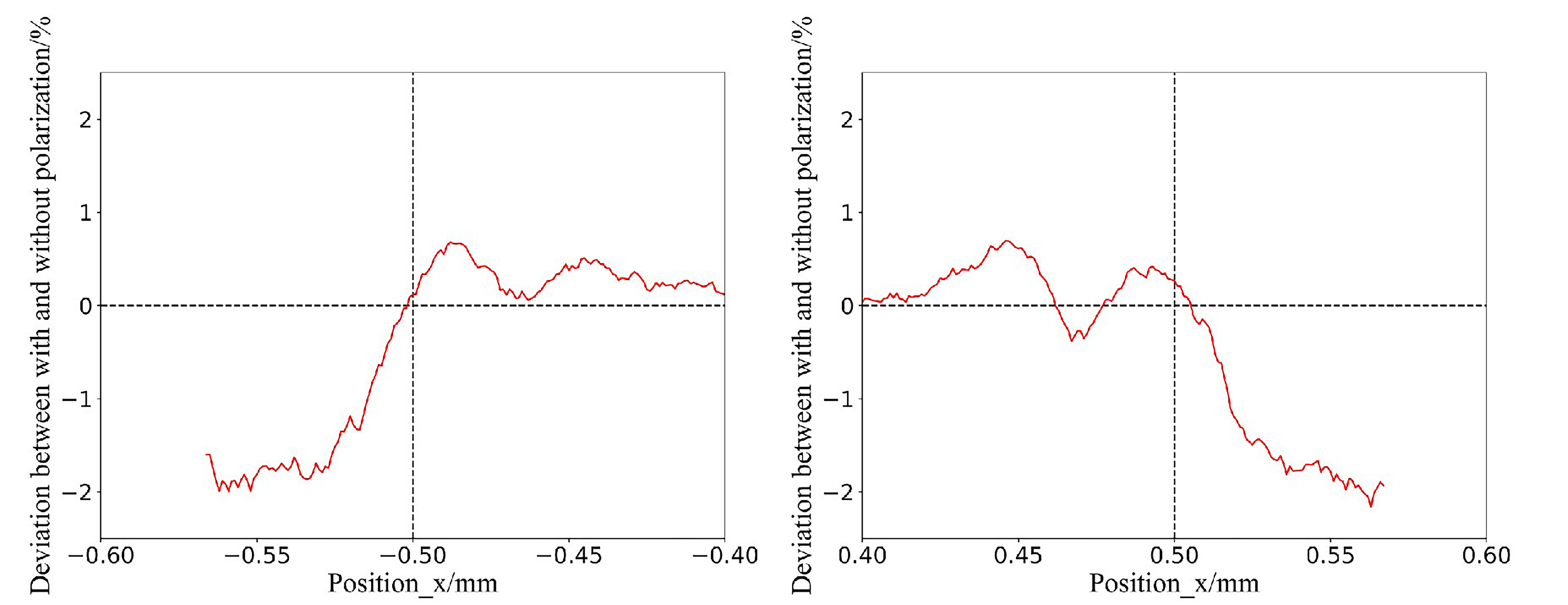}
	\caption{Differences at the edge of the slit between dose calculations regarding and ignoring photon polarization without filters. X-axis represents the horizontal position. Y-axis represents deviation between polarized and non-polarized simulation.}
\end{figure}

In addition to the excellent beam quality, CEPC has a very high dose rate. Such a high dose rate is of great help in FLASH radiotherapy research. The dose rate $Dr$ is calculated using the function as
$$ Dr = \frac{e \times 1.6\times 10^{-19}}{V\times \rho} \times N $$
where $e$ is the deposition energy of a single photon in the selected volume in eV; $V\times \rho$ is the mass of the selected volume in kg; $N$ is the number of photons emitted per second by bending magnet, and $N = 7.19\times 10^{17}$ photons/s.

Finally, through simulation and calculation, the dose rate of the simulated beamline of CEPC has been obtained. Without filters, the average energy of the beam is 134 keV and the dose rate is $1.06 \times 10^{7}$ Gy/s. With filters, the average energy of the beam is 307 keV and the dose rate is $6.13 \times 10^{6}$ Gy/s. In fact, the X-rays produced by CEPC are pulsed structure. According to the specifications provided by the CEPC, the storage ring contains 242 bunches with a revolution frequency of 3003 Hz. This leads to approximately 726,726 bunch pulses per second, with each pulse delivering about 8.44 Gy with filters, based on our earlier calculations. The period of each pulse is approximately $1.38 \times 10^{-6}$ seconds, and the bunch duration is roughly 14.7 picoseconds.\cite{cepc2018cepc} Therefore, the instantaneous dose rate with filters within one pulse is about $5.74 \times 10^{12}$ Gy/s.

\subsection{Prediction model}

A physicochemical model has been developed to briefly predict the treatment effect of FLASH radiotherapy using CEPC synchrotron radiation. With conventional radiotherapy, the effect of lower dose rate on NTCP is small and negligible. However, the influence of ultra-high dose rate on NTCP in FLASH radiotherapy cannot be ignored. Therefore, the physicochemical model is used to relate NTCP to dose and dose rate. In this case, we assume that the beam is continuous photon.

With the mechanism of peroxyl radical $ROO\cdot$ damage to lipids and DNA in normal tissues, correlating the total amount of $ROO\cdot$ exposure in cells during radiation with NTCP is considered. After normalization, the total amount of peroxyl radical $ROO\cdot$ $N(D, Dr)$ can be expressed by the following equation, which is determined by both dose($D$) and dose rate($Dr$).\cite{labarbe2020physicochemical}
\begin{equation}
N(D,Dr) = a(D)+\frac{b(D)}{\sqrt{Dr}}
\end{equation}

where a(D) and b(D) are fitted polynomial coefficients.
By solving the 9 differential rate equations, the corresponding $N(D, Dr)$ of different doses and dose rates can be obtained, and the specific functional forms of a(D) and b(D) are determined by fitting.
\begin{equation}
a(D)=0.09367 \times D^{0.4964}-0.09919
\end{equation}
\begin{equation}
b(D)=0.04214 \times e^{-0.1331 \times D}+0.005073 \times D -0.04196
\end{equation}

Then, for semi-quantitative predictions, the cell biological response and the NTCP can be assumed to be a sigmoid function\cite{labarbe2020physicochemical}
\begin{equation}
NTCP(D,Dr)= (1+exp(-\gamma \times (N(D,Dr)-m)))^{-1}
\end{equation}

where $\gamma$ and m are two constants that can be fitted.

13 groups of experimental data were selected from 5 papers\cite{montay2017irradiation}\cite{alaghband2020neuroprotection}\cite{montay2018x}\cite{montay2021hypofractionated}\cite{vozenin2019advantage} to fit the model, and the parameters $\gamma$ = 1320 and m = 0.2002 was estimated. The side of the function image is shown in the Fig. 6. It can be seen that within a certain range, as the dose rate increases, the tolerated dose of normal tissues also increases.

\begin{figure}
	\includegraphics[width=1\textwidth]{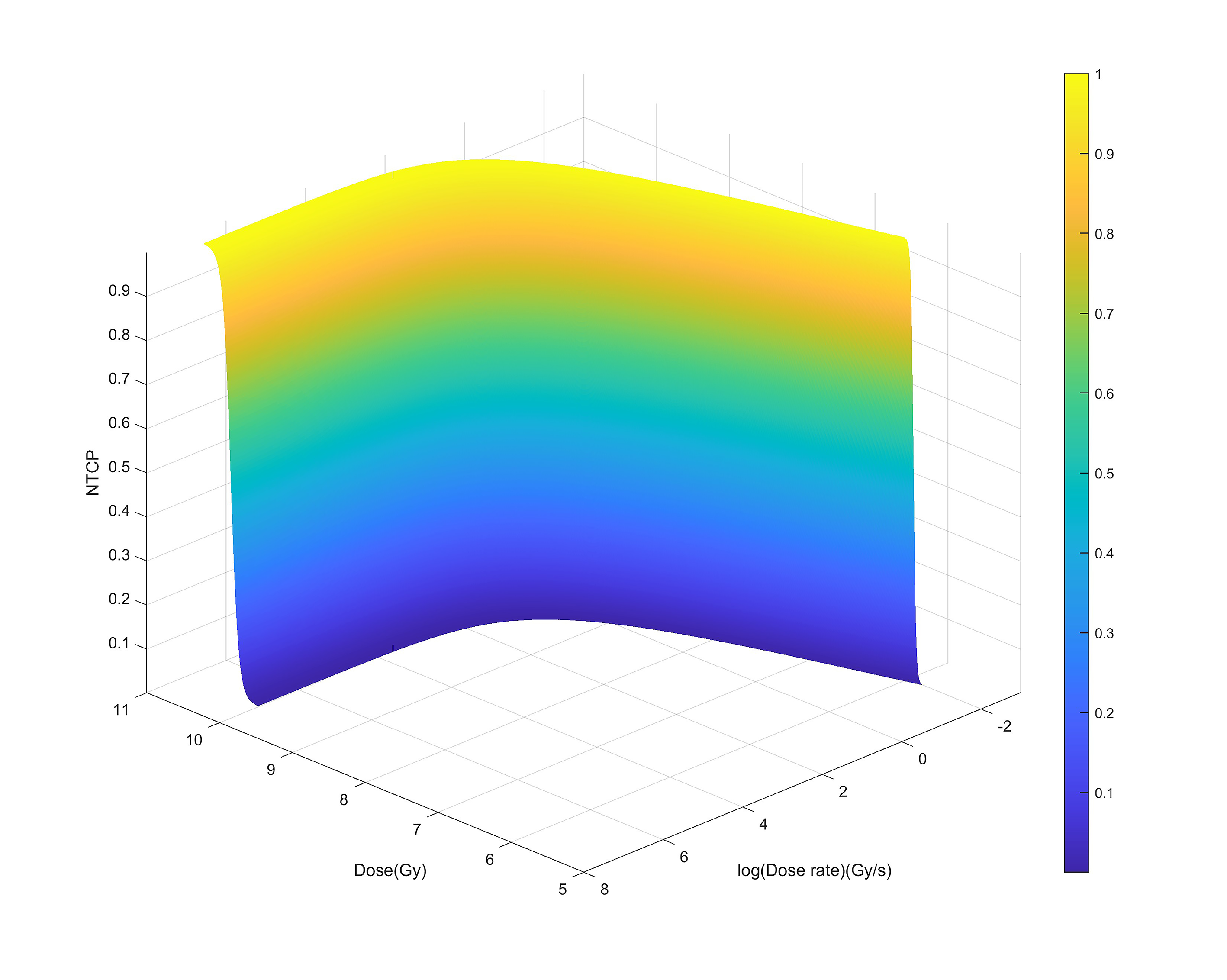}
	\caption{The relationship between NTCP, dose and dose rate. Within a certain range, as the dose rate increases, the tolerated dose of normal tissues also increases.}
\end{figure}

When dose is 10 Gy, the relationship between NTCP and dose rate is shown in the Fig. 7. The 13 black dots represent published experimental data, while the left red dot represents the simulation result of CEPC with filters and the right red dot represents the simulation result of CEPC without filters. The blue shaded area indicates the error margin of the fitness. The X-axis is the dose rate and the Y-axis is NTCP. As can be seen from the figure, FLASH radiotherapy using the medical beamline of CEPC can achieve a good treatment effect.

\begin{figure}
	\includegraphics[width=1\textwidth]{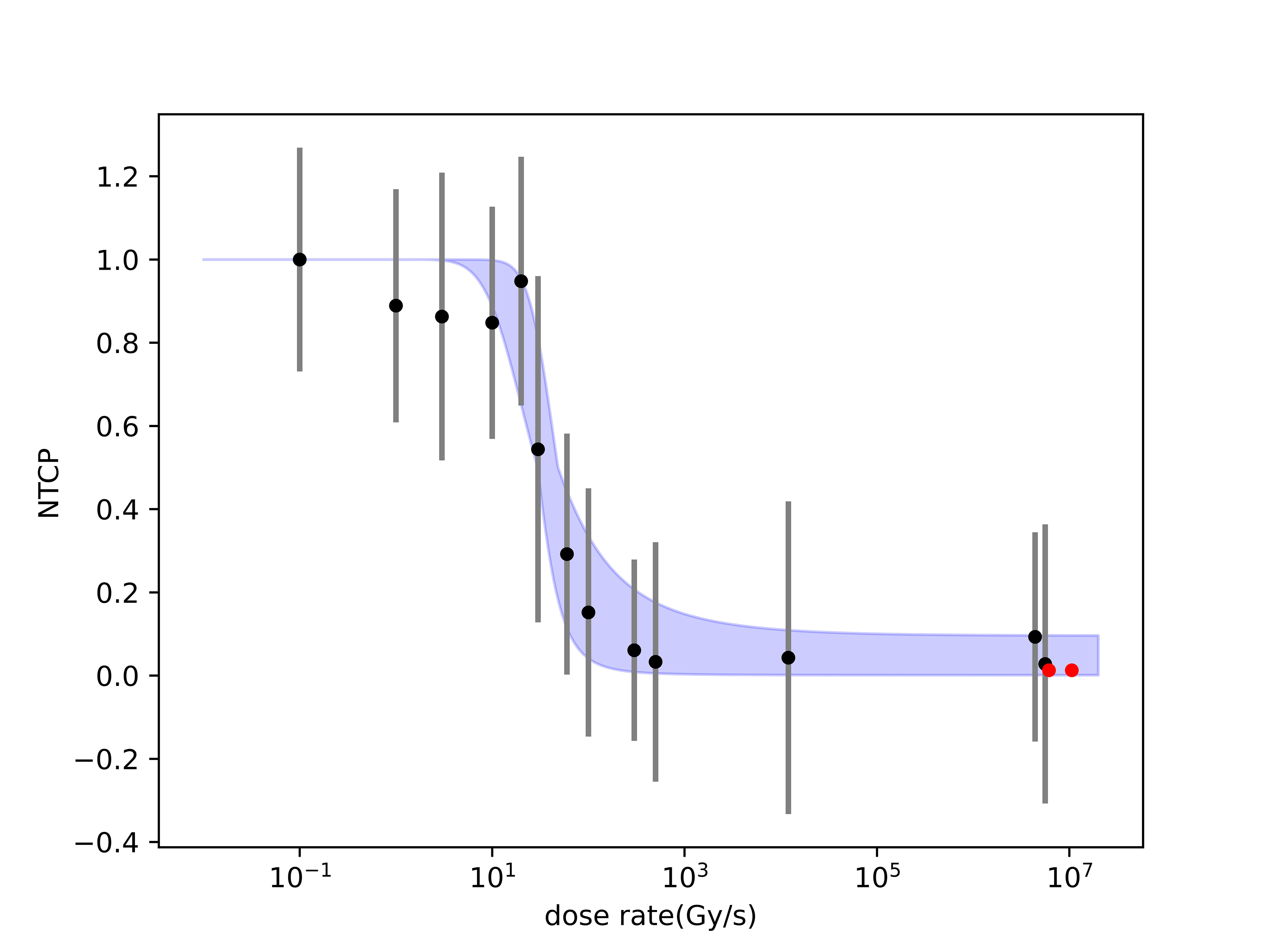}
	\caption{Prediction of treatment effect with CEPC medical beamline. Black dots represent published experimental data, while the left red dot represents the simulation result of CEPC with filters and the right red dot represents the simulation result of CEPC without filters.}
\end{figure}

In our current model, we approximated the beam as continuous rather than considering its pulsed structure. While the work by José Manuel Udías et al. considers the impact of the pulse structure, their findings indicate that the effect of pulsing is minor compared to continuous photon. For example, the free radical ROO· production of pulsed photon is reduced by about 3\% compared to continuous photon.\cite{espinosa2022radical} Therefore, we did not incorporate the pulse structure into our current work. However, we plan to refine the beamline design in future studies, precisely control the dose per pulse, and incorporate the bunch structure into our radio-kinetic model calculations.

\section{Conclusion}

CEPC can generate high-quality synchrotron radiation as a powerful and excellent synchrotron light source, which has great advantages in the medical field. By using SHADOW and Geant4, we successfully built a sample medical beamline of CEPC, simulated the characteristics of the beam emitted by CEPC synchrotron radiation source, and calculated the average energy and the dose rate of the beam. Without filters, the average energy of the beam is 134 keV and the dose rate is $1.06 \times 10^{7}$ Gy/s. With filters, the average energy of the beam is 307 keV and the dose rate is $6.13 \times 10^{6}$ Gy/s.

Then, referring to the physicochemical model of reaction kinetics published by Rudi Labarbe et al., the functional relationship between treatment effect, dose and dose rate was determined by fitting the experimental data of radiotherapy. Finally, the model was used to predict the treatment effect of FLASH radiotherapy with synchrotron radiation of CEPC. The results show that CEPC synchrotron radiation is one of the most promising beams for FLASH radiotherapy.

\bibliography{library}
	
\end{document}